\documentclass{optica-article}

\journal{opticajournal} 

\articletype{Research Article}

\usepackage{lineno}
\usepackage{physics}
\usepackage{siunitx}
\begin{document}

\title{Superheterodyne Rydberg S-band receiver with a multi-tone local oscillator based on an atomic transition loop}

\author{Jan Nowosielski,\authormark{1,2} Mateusz Mazelanik,\authormark{1} Wojciech Wasilewski,\authormark{1,2} and Michał Parniak\authormark{1,2,*}}

\address{\authormark{1}Centre for Quantum Optical Technologies, Centre of New Technologies, University of Warsaw, S. Banacha 2c, 02-097 Warsaw, Poland\\
\authormark{2}Faculty of Physics, University of Warsaw, L. Pasteura 5, 02-093 Warsaw, Poland\\}

\email{\authormark{*}mparniak@fuw.edu.pl} 


\begin{abstract*} 
Atomic-vapor sensors based on Rydberg atoms now face a transition towards practical applications, with several outstanding challenges. To achieve the best sensitivities, a superheterodyne mode of operation is desired, which requires the presence of a local oscillator in the vapor cell. This local oscillator hinders several advantages of the sensor, such as stealthy and all-optical operation. We propose and realize a detection scheme that avoids some of those problems by using multi-tone mixing, where direct usage of the local oscillator at the same frequency is not required. Our scheme is further elaborated on using efficient theoretical methods to predict the performance of the sensor. Our sensor operates at the S-band frequency, known for its usage in IEEE 802.11 (Wi-Fi) networks, without interfering with the signal itself.
\end{abstract*}

\section{Introduction}





Rydberg atom-based electromagnetic (EM) radiation sensors are a promising technology already holding a solid position in the scientific community. The technology utilises large transition dipole moments between Rydberg states to achieve EM field sensitivity on par with specialized electronics \cite{Fancher2021,SantamariaBotello2022}. Yet, it still allows for high detection bandwidth \cite{Meyer2021, Cui2023} and offers advantages such as extraordinary tunability and, importantly, self calibration feature based on fundamental constants. These naturally arose features, combined with the possibility of relatively low complexity of the sensors and the possibility of miniaturization and packaging \cite{Simons2018, Mao2023, Zhao2023} drove the attention of the engineering community that joined the effort of maturing the technology. As technology matures, new scientific and practical applications are born.

The established detection schemes allow measuring the RF EM field amplitude \cite{Sedlacek2012, Osterwalder1999}, detecting frequency and amplitude modulation \cite{Borowka2022, Liu2021}, phase-sensitive measurement \cite{Simons2019, Gordon2019, Jing2020}, and photon counting via RF-to-optical conversion \cite{Borowka2024a, Han2018, Vogt2019}. Moreover, the properties of Rydberg atoms allow for measurement of different properties of the oncoming EM field, such as angle of arrival \cite{Robinson2021}, and polarization \cite{Sedlacek2013,Anderson2018}. Apart from applications in electrometry, in recent years, multiple groups have presented communication schemes employing phase-sensitive Rydberg-based receivers \cite{Meyer2018, Song2019}.

The first sensors to be field-deployed are most likely the simplest ones. The lowest complexity, amplitude sensor based on Autler-Townes (AT) effect spectroscopy, while having limited sensitivity, enables self-referencing by linking the measured amplitude with the Rabi frequency and thus dipole moment. However, in many applications, higher sensitivity and phase-resolving detection are desired. This is offered by the superheterodyne sensor that employs an additional field acting as a local oscillator which as in the case of standard phase-sensitive detection of the electronic signal, needs to be close in frequency and stronger than the signal itself. Thus, the additional field can disturb the source of the measured field and is often generally undesired. 
Interestingly enough, phase-sensitive detection can be achieved without employing a local oscillator utilizing the loop energy schemes. The phase-sensitivity of such setups has been described theoretically \cite{Morigi2002, Kosachiov1992, Buckle1986}, and in recent years, multiple groups have utilized it to perform the phase-sensitive detection in case of all-optical \cite{Berweger2023a} and microwave-optical \cite{Borowka2024, Berweger2023a, Anderson2022} schemes. An interesting case arises when considering a fully microwave loop, which up to this moment was proposed only theoretically \cite{Shylla2018}, as the phase between generated microwave fields can be easily controlled electronically. In this paper, we propose an experimental implementation of such a setup consisting of three different microwave fields, and compare measured phase-dependent probe transmission spectra as fields to the numerical predictions. Additionally, we characterize our setup as a receiver, finding its sensitivity and response range.

\section{Experimental setup}
\subsection{Energy level scheme}
In the experiment, we consider a 6-level energy ladder of $^{85}\mathrm{Rb}$ depicted in the Fig \ref{fig:setup}a. In the following setup, the probe laser is tuned to the $\mathrm{D}_2$ transition between ground state $5^2\mathrm{S}_{1/2}(F=3)$ and $5^2\mathrm{P}_{3/2}(F=4)$. The second $\SI{776}{\nm}$ and third $\SI{1268}{nm}$ fields coupled to the $5^2\mathrm{P}_{3/2}(F=4)\rightarrow5^2\mathrm{D}_{5/2}(F=5)$ and $5^2\mathrm{D}_{5/2}(F=5)\rightarrow32^2\mathrm{F}_{7/2}$ transitions respectively excite atoms to the Rydberg state. The overlapping electromagnetically induced transparency effects (EIT) \cite{Fleischhauer2005} caused by both fields lead to the emergence of the so-called electromagnetically induced absorption effect (EIA) \cite{Carr:12}.

Next field, coupled to the $32^2\mathrm{F}_{7/2}\rightarrow32^2\mathrm{G}_{9/2}$ transition with the frequency, we call the signal ($\mathsf{SIG}$) microwave ($\mathsf{MW}$) field with the frequency $f_{\mathsf{SIG}}\approx \SI{2.5}{\GHz}$. To perform the phase-sensitive measurement of the $\mathsf{SIG}$ field, we utilize the all-microwave loop interferometry scheme with two additional $\mathsf{MW}$ fields. The dressing ($\mathsf{DRS}$) field with the frequency $f_{\mathsf{DRS}} \approx \SI{500}{\MHz}$ excites atoms through the $32^2\mathrm{G}_{9/2}\rightarrow32^2\mathrm{H}_{11/2}$ transition and the loop is then closed by the coupling ($\mathsf{CPL}$) field at the frequency $f_{\mathsf{CPL}}\approx \SI{1.5}{\GHz}$, which drives atoms from the $32^2\mathrm{F}_{7/2}$ to the $32^2\mathrm{H}_{11/2}$ via the two-photon transition.

It was shown in other works that in the case of the resonance of all the fields in the loop, i.e. $f_{\mathsf{SIG}}+f_{\mathsf{DRS}}-2f_{\mathsf{CPL}}=0$ the probe transmission through the atomic medium depends solely on the phase between fields. In the non-resonant case, that is if one of the fields is detuned from the loop resonance, the phase between fields becomes time-dependent and is equal $\varphi = 2\pi\cdot f_{\mathsf{OPT}} t$, where $t$ is time and $f_{\mathsf{OPT}} = f_{\mathsf{SIG}}+f_{\mathsf{DRS}}-2f_{\mathsf{CPL}}$ is the frequency mismatch, which can be interpreted as the frequency of the optical ($\mathsf{OPT}$) signal. Moreover, if the $\mathsf{SIG}$ was phase-modulated, the modulation could then be recovered by measuring the transmission via the photodiode ($\mathsf{PD}$) and then demodulating the measured $\mathsf{PD}$ signal at the beat note frequency.

\subsection{Experimental setup}
\begin{figure}
   \centering
  \includegraphics[width=11cm]{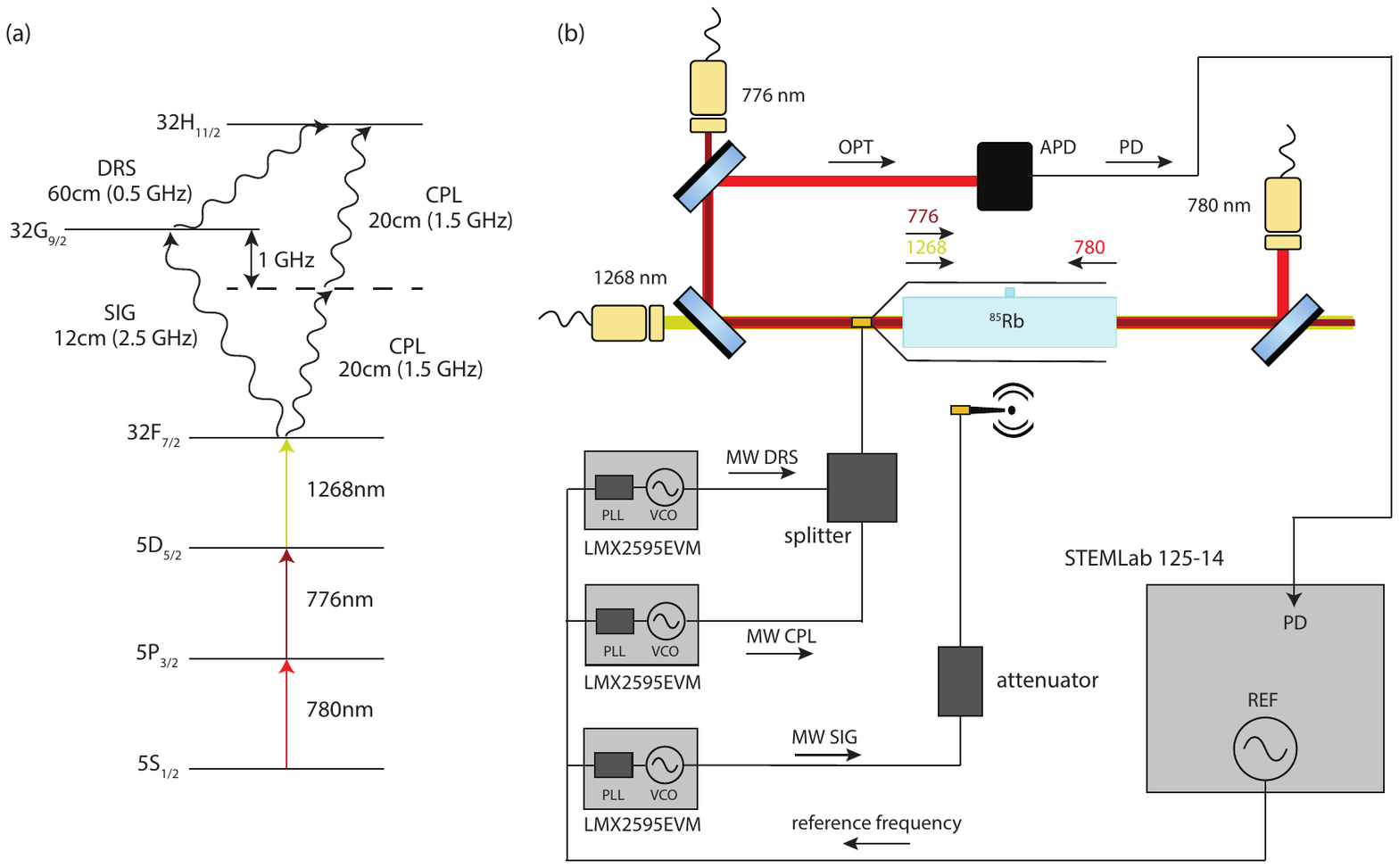}
\caption{(a) $^{85}\mathrm{Rb}$ energy level configuration used in the experiment. (b) Scheme of the experimental setup used. $\mathsf{MW}$ - microwave, $\mathsf{SIG}$ - signal, $\mathsf{CPL}$ - coupling, $\mathsf{DRS}$ - dressing, $\mathsf{APD}$ - avalanche photodiode, $\mathsf{OPT}$ - optical signal, $\mathsf{PD}$ - photodiode signal, $\mathsf{PLL}$ - phase-locked loop, $\mathsf{VCO}$ – voltage-controlled oscillator.}
\label{fig:setup}
\end{figure}

Our experimental setup is built around room-temperature ($\SI{22.5}{\celsius}$) $^{85}\mathrm{Rb}$ atoms, and its simplified scheme is depicted in Fig \ref{fig:setup}b. To partially reduce the thermal (Doppler) broadening, the pair of 776 and 1268 nm beams and the probe beam counter-propagate in the cell. All beams are circularly polarized, with the probe beam being left-handed circularly polarized and coupling beams being right-handed circularly polarized. All beams are focused inside the cylindrical vapor cell with a length equal to 58 mm and beams waist size equal to $w = \SI{300}{\micro\meter}$.. The powers of lasers are chosen to maximize the EIA effect, which is achieved for probe laser power equal to $P_{780} =\SI{1.5}{\micro\watt}$, and powers of 776 and 1268 nm lasers equal to $P_{776} = \SI{2.4}{\milli\watt}$ and $P_{1268} = \SI{85}{\milli\watt}$. All lasers are frequency-stabilized to the master laser via the cavity setups.

The $\mathsf{MW}$ signals are generated using the LMX2595EVM frequency synthesizers. The $\mathsf{SIG}$ field is generated by sending a signal at the frequency $f_{\mathsf{SIG}}=\SI{2514}{\MHz}$ via the controllable attenuator to the commercially available Wi-Fi antenna. The $\mathsf{CPL}$ and $\mathsf{DRS}$ fields are generated by two additional frequency synthesizers at the frequencies $f_{\mathsf{DRS}} =\SI{510.4}{\MHz}$ and $f_{\mathsf{CPL}}=\SI{1512.2}{\MHz}$. The signals are then combined via the frequency splitter and sent to the capacitor-like antenna mounted around the rubidium cell. The frequencies of the $\mathsf{MW}$ fields are chosen to optimize the atomic response. To have a stable and controllable phase between the $\mathsf{MW}$ fields, all the frequency synthesizers get the same reference frequency from the STEMLab 125-14 multipurpose tool.

To perform the phase-sensitive detection, used to calibrate the receiver, as well as measure atomic response range, the 776 and 1268 nm lasers are tuned to their respective transitions and the probe laser is detuned by $\SI{3}{\MHz}$ to the lower frequencies to maximize the atomic response to the $\mathsf{MW}$ fields. The probe transmission is measured by the avalanche photodiode with the $\SI{50}{\MHz}$ bandwidth, and the measured $\mathsf{PD}$ signal is sent to the STEMLab 125-14 multipurpose tool for further processing. Powers of the $\mathsf{CPL}$ and $\mathsf{DRS}$ fields for the measurements of sensitivity and atomic response range are also optimized to maximize the atomic response and the Rabi frequencies for both transitions are equal to $\Omega_{\mathsf{DRS}} = 2\pi\cdot\SI{10.1}{\MHz}$ and $\Omega_{\mathsf{CPL}} = 2\pi\cdot\SI{7.5}{\MHz}$. It is important to note that the mentioned $\Omega_{\mathsf{CPL}}$ refers to the effective Rabi frequency of the two-photon transition.
\section{Results}
\subsection{Comparison with theoretical predictions}
As mentioned in a previous section, the loop scheme consisting only of the $\mathsf{MW}$ fields allows for full control and prediction of the phase between the fields. For that reason, it is possible to fit the theoretical model predicting the behaviour of the probe transmission as a function of phase between $\mathsf{MW}$ fields to the experimental data. The experimental data was gathered for the case, where the EIT caused by $\SI{1268}{\nm}$ laser was detuned by about $\SI{30}{\MHz}$ to the lower probe frequencies, so the EIT caused by $\SI{776}{\nm}$ and $\SI{1268}{\nm}$ lasers do not overlap. Moreover, to get better fits of the transmission spectra, the $\mathsf{SIG}$ and $\mathsf{DRS}$ were detuned by $\Delta = -\SI{5}{\MHz}$. The measurements are performed by measuring the probe transmission with a locked probe laser. After each measurement, the probe laser is relocked at a frequency detuned by $\Delta f$ from the previous one, and the probe transmission is measured again. The sequence is then repeated over the probe frequency range, where the EIT caused by $\SI{1268}{\nm}$ laser is visible.

The numerical predictions of the probe laser transmission were found in all the cases by solving the so-called master equation:
\begin{equation}
    \dot{\rho}(t) = \mathcal{L}\rho(t),
\end{equation}\label{eq:master}
where $\rho$ is the density matrix of the considered atomic level configuration and $\mathcal{L}$ is a Lindblad superoperator describing the evolution of the system. Including the decoherence due to the natural lifetime of the excited states, it can be written as:
\begin{equation}
    \mathcal{L}\rho =  -\frac{i}{\hbar}(H_c\rho - \rho H_c^\dagger) + \mathcal{R}\rho
\end{equation}
where $\mathcal{R}$ is the repopulation operator and $H_c = H - \frac{i}{2}\sum_\beta \Gamma_\beta \ket{\beta}\bra{\beta}$ is the conditional Hamiltonian of the setup described by Hamiltonian $H$, taking into account decoherence rates $\Gamma_\beta$. In case of the full energy levels scheme, the conditional Hamiltonian is given as:
\begin{equation}
    H_c = -\frac{1}{2}\begin{pmatrix}
0 & \Omega^*_{780} & 0 & 0 & 0 & 0 \\
\Omega_{780} & 2\Delta_{\textbf{I}}+i\Gamma_1 & \Omega^*_{776} & 0 & 0 & 0 \\
0 & \Omega_{776} & 2\Delta_{\textbf{II}}+i\Gamma_2 & \Omega^*_{1268} & 0 & 0 \\
0 & 0 & \Omega_{1268} & 2\Delta_{\textbf{III}}+i\Gamma_3 & \Omega^*_{\mathsf{SIG}} & \Omega^*_{\mathsf{CPL}} \\
0 & 0 & 0 & \Omega_{\mathsf{SIG}} & 2\Delta_{\textbf{IV}}+i\Gamma_4 & \Omega^*_{\mathsf{DRS}}e^{-i\varphi(t)} \\
0 & 0& 0 & \Omega_{\mathsf{CPL}} & \Omega_{\mathsf{DRS}}e^{i\varphi(t)} & 2\Delta_{\textbf{V}}+i\Gamma_5
    \end{pmatrix}
\end{equation}
Here, the $\Omega_n$ are Rabi frequencies of the corresponding fields, $\Gamma_n$ are decoherence rates of the $n$--th state and $\varphi(t)$ is the time-varying loop phase between microwave fields. The detunings from energy levels $\Delta$ are defined as:
\begin{equation}
    \begin{split}
        \Delta_{\textbf{I}} &= \Delta_{780} \\
        \Delta_{\textbf{II}}&= \Delta_{780} + \Delta_{776} \\
        \Delta_{\textbf{III}} &= \Delta_{780} + \Delta_{776} + \Delta_{1268}\\
        \Delta_{\textbf{IV}} &= 
        \Delta_{780} + \Delta_{776} + \Delta_{1268} + \Delta_{\mathsf{SIG}}\\
        \Delta_{\textbf{V}} &= 
        \Delta_{780} + \Delta_{776} + \Delta_{1268} + 2\Delta_{\mathsf{CPL}}
    \end{split}
\end{equation}

For computational simplicity, the two-photon transition was assumed to be a single transition between two states. Thus, the Rabi frequency of the $\mathsf{CPL}$ field is the effective Rabi frequency of the two-photon transition. In the case of the steady state solution, used to determine the Rabi frequencies of the $\mathsf{SIG}$ and $\mathsf{DRS}$ fields, we assume that both the density matrix and Hamiltonian of the system are independent of time, thus reducing Eq.~\ref{eq:master} to the form of an easily computable $\mathcal{L}\rho = 0$. However, when considering the time-dependent equation, as in the case of the transition loop, the differential equation needs to be solved. Additionally, the Doppler broadening effect is taken into account by averaging the found density matrix over the velocity distribution.  A more detailed and thorough explanation of the theoretical model, as well as the numerical methods used to determine the solution of the master equation, can be found in a recent paper from our group \cite{Kasza2024}.

As the impact of the $\mathsf{SIG}$ and $\mathsf{DRS}$ could be directly seen in the probe transmission spectrum, the Rabi frequencies of both those fields were found beforehand by fitting the transmission spectra to the steady state solution of the Lindblad equation for this path of the energy level loop. The effective Rabi frequency of the two-photon transition induced by the $\mathsf{CPL}$ was then found by fitting the solution of the Lindblad equation to the resonant case of the closed loop with all the $\mathsf{MW}$ fields turned on. Using the found values, we performed the calculations for the time-dependent case. We compare the theoretical predictions with the experimental data in two situations: one called weak field regime, where the Stark shift of the overlapping Rydberg states due to the $\mathsf{MW}$ fields is negligible, and one in the strong field regime, where it becomes prominent and impacts the quality of the fit.

\begin{figure}
  \centering
  \includegraphics[width=11cm]{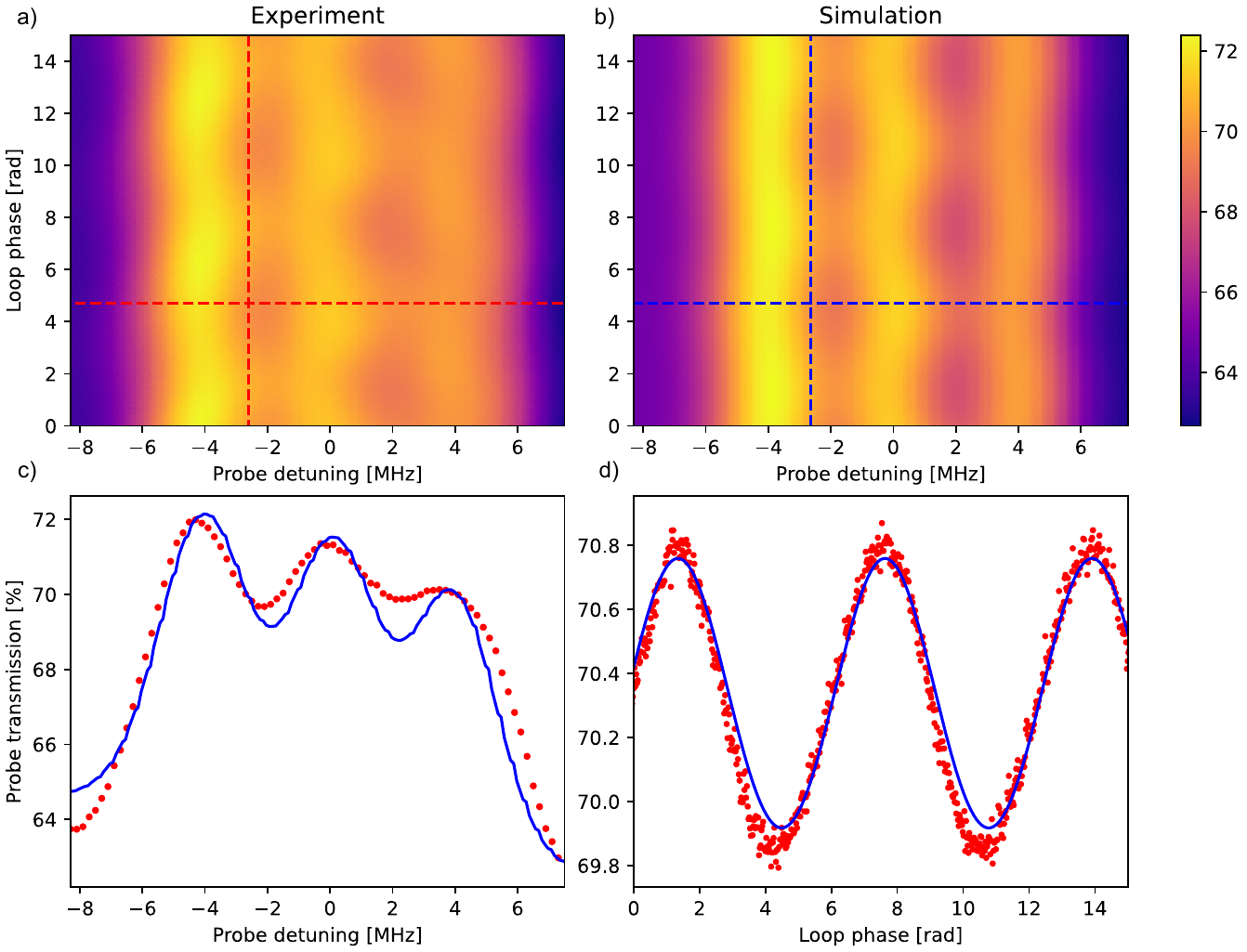}
\caption{Comparison between measured data and theoretical predictions in the weak field case. a) measured probe transmission as a function of probe detuning and the phase between $\mathsf{MW}$ fields. b) numerical predictions of the probe transmission as a function of probe detuning and the phase between $\mathsf{MW}$ fields. c) cross-section through the specific phase, dashed line represents experimental data, and solid line represents theoretical predictions. d) cross-section through the probe detuning yielding the strongest response to the $\mathsf{MW}$ fields, dashed line represents experimental data, and solid line represents theoretical predictions.
}
\label{fig:compare_low}
\end{figure}

Both measurements were performed for the beat note frequency equal to $f_{\mathsf{OPT}} = \SI{1.9}{\kHz}$. In the case of the weak field regime, the Rabi frequencies of the $\mathsf{MW}$ fields were equal to $\Omega_{\mathsf{SIG}} = 2\pi\cdot\SI{10.5}{\MHz}$, $\Omega_{\mathsf{DRS}} = 2\pi\cdot\SI{7.8}{\MHz}$ and $\Omega_{\mathsf{CPL}} = 2\pi\cdot\SI{0.5}{\MHz}$. The comparison between the experimental data and the theoretical predictions is shown in Fig.~\ref{fig:compare_low}. The 2D maps in the upper row represent the probe transmission as a function of the phase between $\mathsf{MW}$ fields and the probe detuning for the experimental data seen in Fig.~\ref{fig:compare_low}a and simulations Fig.~\ref{fig:compare_low}b. The dashed lines on both maps represent the cross-sections shown in the lower row, with the cross-section through the phase seen in Fig.~\ref{fig:compare_low}c and the cross-section through the probe detuning yielding the strongest $\mathsf{MW}$ response seen in Fig.~\ref{fig:compare_low}d. It can be noted that in the weak fields regime, the theoretical prediction fits experimental data properly, and the behaviour of the probe transmission as a function of the loop phase can be predicted.

\begin{figure}
  \centering
  \includegraphics[width=11cm]{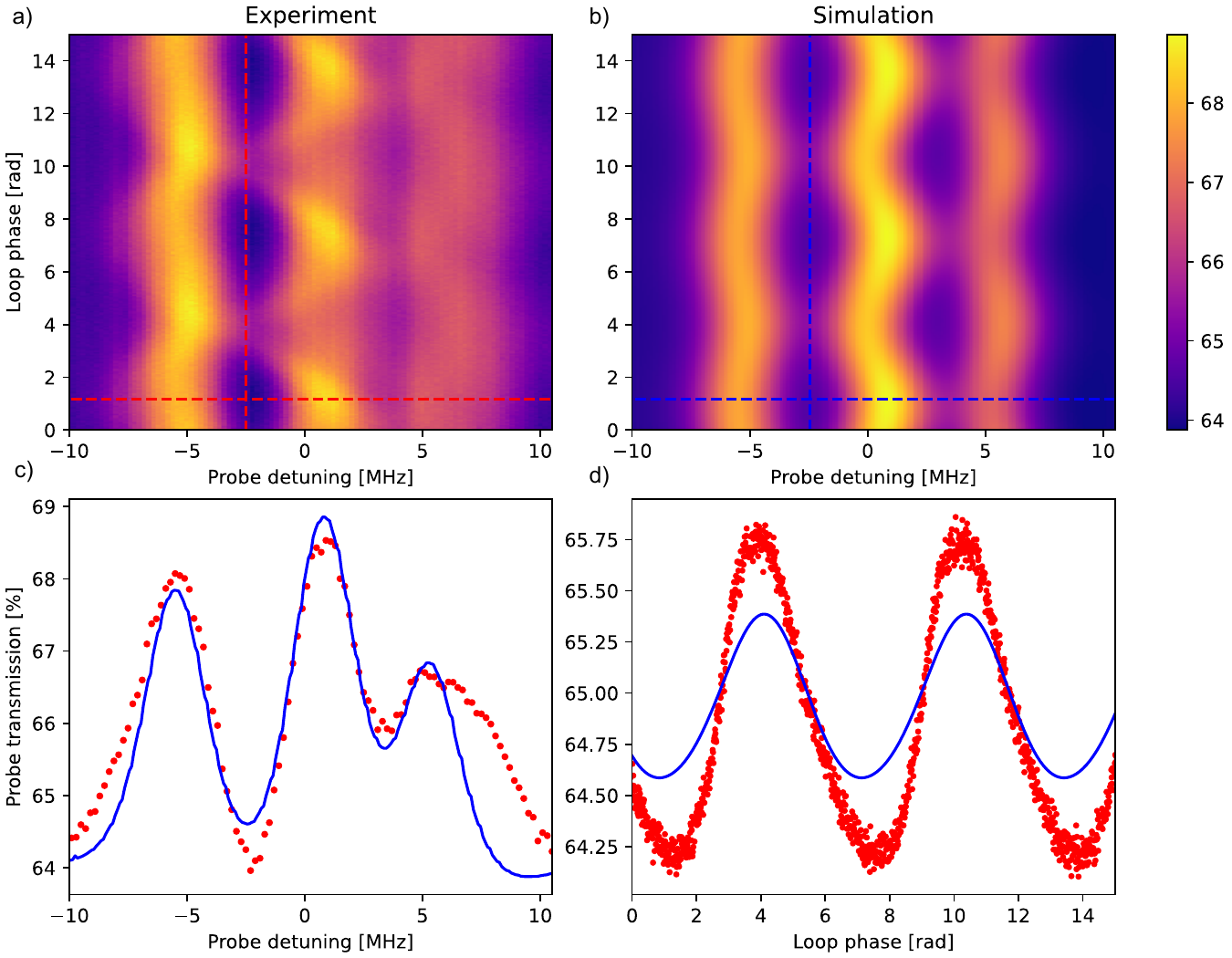}
\caption{Comparison between measured data and theoretical predictions in the strong field case. a) measured probe transmission as a function of probe detuning and the phase between $\mathsf{MW}$ fields. b) numerical predictions of the probe transmission as a function of probe detuning and the phase between $\mathsf{MW}$ fields. c) cross-section through the specific phase, dashed line represents experimental data, and solid line represents theoretical predictions. d) cross-section through the probe detuning yielding the strongest response to the $\mathsf{MW}$ fields, dashed line represents experimental data, and solid line represents theoretical predictions.
}
\label{fig:compare_high}
\end{figure}

In the case of strong fields, the Rabi frequencies were found to be $\Omega_{\mathsf{SIG}} = 2\pi\cdot\SI{14}{\MHz}$, $\Omega_{\mathsf{DRS}} = 2\pi\cdot\SI{10.1}{\MHz}$ and $\Omega_{\mathsf{CPL}} = 2\pi\cdot\SI{1.8}{\MHz}$. The comparison between the experimental data and numerical predictions is shown in Fig.~\ref{fig:compare_high}. The mismatch between theoretical predictions and experimental data can be seen, which becomes more prominent for the positive probe detunings. Such divergence from the numerical simulations can be attributed to the Stark shifts of the degenerate Rydberg states, caused mostly by the single-photon effects caused by the $\mathsf{CPL}$ field. As the transition dipole moments between degenerate hyperfine states of the Rydberg levels differs, the induced Stark shift varies between each of the transitions and, for weaker fields, can be seen in the transmission spectrum as broadening of the EIT peaks.
\subsection{Calibration of the microwave field}
The whole calibration procedure uses similar methods as our previous work regarding $\mathsf{MW}$ field detection at these transitions \cite{Nowosielski2024}. To find the receiver's sensitivity and the noise level, the calibration of its response to the $\mathsf{MW}$ fields is required. For the calibration, we detune the $\SI{1268}{\nm}$ laser by $\SI{25}{\MHz}$ to the lower probe frequencies, so the EIT caused by the $\SI{1268}{\nm}$ and $\SI{776}{\nm}$ do not overlap with each other. We then measure the transmission spectrum with the EIT splitting caused by the $\mathsf{SIG}$ field. The measurement is performed for the highest power of the $\mathsf{SIG}$ field being $\SI{-15}{\decibel m}$. To change the power of the $\mathsf{SIG}$ field we use the controllable attenuator precise up to $\pm \SI{0.5}{\decibel m}$ and an additional $\SI{3}{\decibel m}$ and $\SI{10}{\decibel m}$ attenuators which were added between measurements to probe the full dynamic range of the receiver. For the absolute calibration, we measure the transmission spectra for the attenuations from 0 to 15 dB with the step of 3 dB. To perform the absolute calibration, we then fit the numerical solutions for the steady state of the Linblad equation to the experimental data, from which we find the Rabi frequencies of the $\mathsf{SIG}$ field as a function of the signal attenuation. Based on the found Rabi frequencies, we calculate the amplitude of the electric field given as $A=\frac{\hbar \Omega}{d}$, where $\Omega$ is the Rabi frequency and $d$ is the dipole moment of the transition found using the Alkali Rydberg Calculator \cite{Robertson2021}. In the considered experimental setup, the dipole moment equals $d = 384a_0 e$, where $a_0$ is the Bohr radius and $e$ is the electron charge. To find the absolute calibration line, we then fit a line function with a fixed slope equal to $-1$ to the electric field amplitude as a function of the signal attenuation in a logarithmic scale, thus acquiring the constant term of the calibration line.

In the case of the weaker $\mathsf{SIG}$ fields, absolute calibration is impossible due to their impact not being visible directly in the absorption spectrum. Because of that, the receiver's response needs to be measured differently, and then calibrations in both regimes can be merged to acquire the relation between the electric field amplitude and the signal attenuation spanning the whole response range. To measure the response in that regime we utilize the phase-sensitive detection with the $\mathsf{OPT}$ signal frequency equal to $f_{\mathsf{OPT}} = \SI{5}{\MHz}$, for which we measure 100 subsequent waveforms of the Fourier power spectra of the $\mathsf{PD}$ signal over the $\SI{0.26}{\ms}$. We repeat that measurement for the $\mathsf{SIG}$ powers of $-15,\,-17,\,-21,\,-34\,\SI{}{\decibel m}$ and signal attenuations from $\SI{0}{\decibel}$ to $\SI{30}{\decibel}$ with the step of $\SI{2}{\decibel}$. Based on measured Fourier power spectra we then find the $\mathsf{PD}$ signal power level by taking the maximum value of the averaged Fourier power spectrum in the range of $\SI{1}{\MHz}$ around the expected value of $f_{\mathsf{OPT}}$. Additionally, the noise power level is found by similarly measuring the noise power spectra and then averaging it over the frequency range in which we looked for the $\mathsf{PD}$ signal power level. 

The calculated values of the $\mathsf{PD}$ signal power level are then considered on a logarithmic scale, in which the power level changes linearly with the attenuation. For each $\mathsf{SIG}$ power, we fit the linear function with the slope value fixed to -1. Points for the highest signal power for the phase-sensitive detection are shifted to lie on the same line as points from the absolute calibration, allowing for calculating the amplitude of the electric field based on the $\mathsf{PD}$ signal power level for the phase-sensitive detection. Fitted lines are then shifted to continue the absolute calibration line,
spanning the whole range of the atomic response from saturation down to the noise level. Moreover, shifting the calibration lines to extend the absolute calibration line gives us the relation between the amplitude of the electric field and the $\mathsf{PD}$ signal power level.

\begin{figure}
  \centering
  \includegraphics[width=11cm]{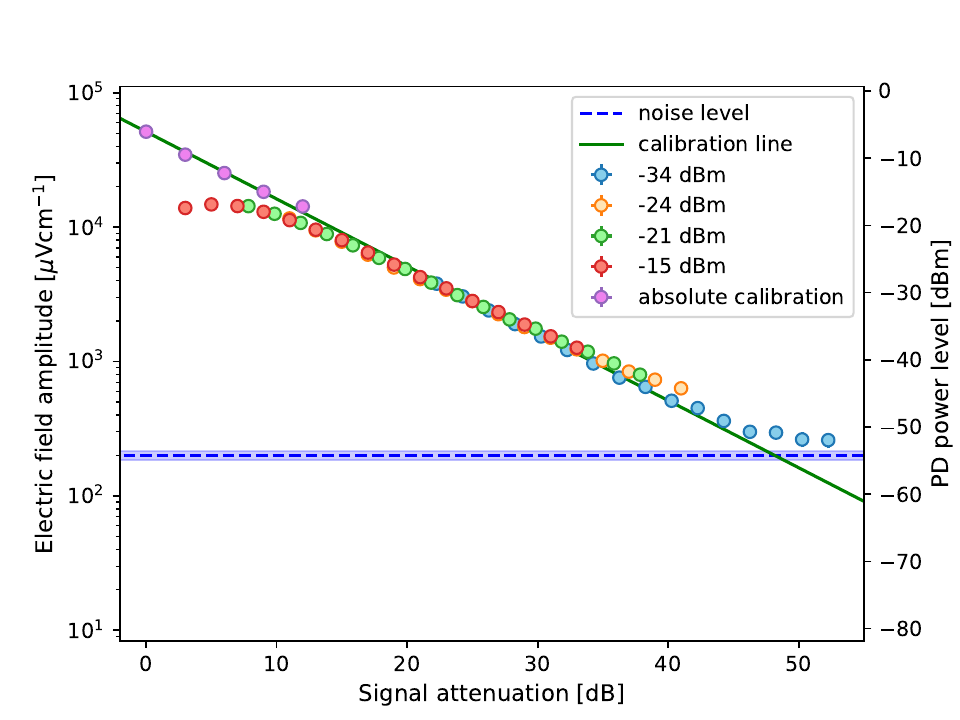}
\caption{Amplitude of the $\mathsf{SIG}$ field (left vertical axis) as a function of signal attenuation. The right vertical axis refers to the $\mathsf{PD}$ signal power level calculated from the Fourier power spectrum for phase-sensitive detection. The data points were gathered during the absolute calibration and phase-sensitive detection. The blue area around the shot noise level corresponds to the standard error of the calculated value.}
\label{fig:calib}
\end{figure}

Using the found relation, we calculate the amplitude of the electric field for which the $\mathsf{PD}$ signal power level is equal to the noise power level, which was checked to be the shot-noise of the probe laser. The electric field amplitude corresponding to that signal level is equal to $\SI{200\, \pm 20}{\micro\volt\per\cm}$ which in case of the $\SI{0.26}{\ms}$ measurement time window translates to a noise level equal to $\SI{3.2\pm0.3}{\micro\volt\per\cm\per\hertz\tothe{0.5}}$. The relation between the electric field amplitude and the $\mathsf{PD}$ signal attenuation together with the calibration line can be seen in Fig \ref{fig:calib}.
We also find the amplitude of the $\mathsf{SIG}$ field corresponding to the saturation level, which is defined as a power level 1 dB weaker than the power threshold for which the atomic response due to changing signal attenuation becomes nonlinear. The calculated value of the amplitude of the electric field corresponding to the saturation level is equal to $A_{sat} = \SI{7.5\pm0.2}{\milli\volt\per\cm}$. 
\subsection{Atomic response range}
\begin{figure}
  \centering
  \includegraphics[width=11cm]{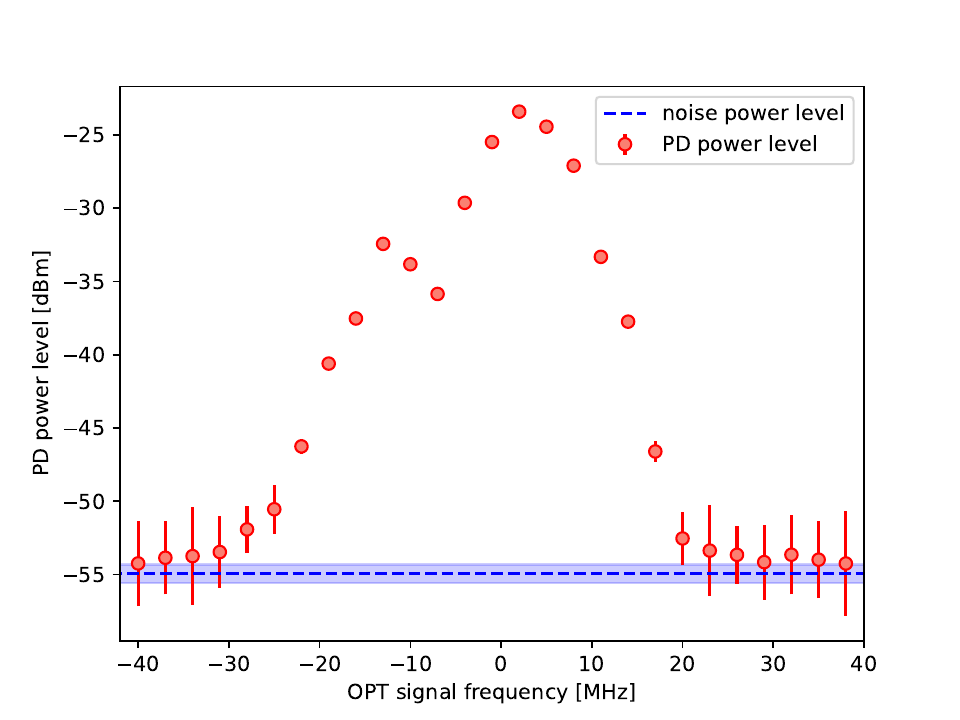}
\caption{Atomic response power level as a function of the frequency of the $\mathsf{OPT}$ signal together with the shot-noise power level. The atomic response was measured for different frequencies of the $\mathsf{SIG}$ field, yielding different values of $f_{\mathsf{OPT}}$. The blue area around the shot-noise power level corresponds to the standard error of the calculated value.}
\label{fig:range}
\end{figure}

To further characterize the receiver, we measure the atomic response range in two distinct cases. In both cases, the amplitude of the $\mathsf{SIG}$ field is constant and equal to $A = \SI{7.0\pm0.2}{\milli\volt\per\cm}$. In the first case, we only change the $f_{\mathsf{SIG}}$, while keeping the rest of the $\mathsf{MW}$ fields at the constant frequencies, thus also changing the $f_{\mathsf{OPT}}$ in the process. To measure the atomic response at different frequencies, we change the $\mathsf{SIG}$ frequency by $\SI{3}{\MHz}$ in the range from $\SI{2480}{\MHz}$ to $\SI{2540}{\MHz}$. For each $\mathsf{SIG}$ frequency, we measure 100 waveforms of the Fourier power spectrum of the $\mathsf{PD}$ signal with the measurement time of $\SI{0.26}{\ms}$. The $\mathsf{PD}$ power level is then found by taking the maximum value of the power spectrum in the range of $\SI{1}{\MHz}$ around the expected value of $f_{\mathsf{OPT}}$. Such measured atomic response as a function of the $\mathsf{SIG}$ field frequency can be seen in Fig \ref{fig:range} together with the shot-noise power level derived by averaging the noise power level over the whole spectrum and all the measurements. As it can be seen in Fig \ref{fig:range}, the atomic response drops to the shot-noise level at the $\mathsf{OPT}$ signal frequencies of about $f_{\mathsf{OPT}} = \SI{\pm 20}{\MHz}$, which translates to the $f_{\mathsf{SIG}} = \SI{2534}{\MHz}$ and $f_{\mathsf{SIG}} = \SI{2494}{\MHz}$. 

\begin{figure}
  \centering
  \includegraphics[width=11cm]{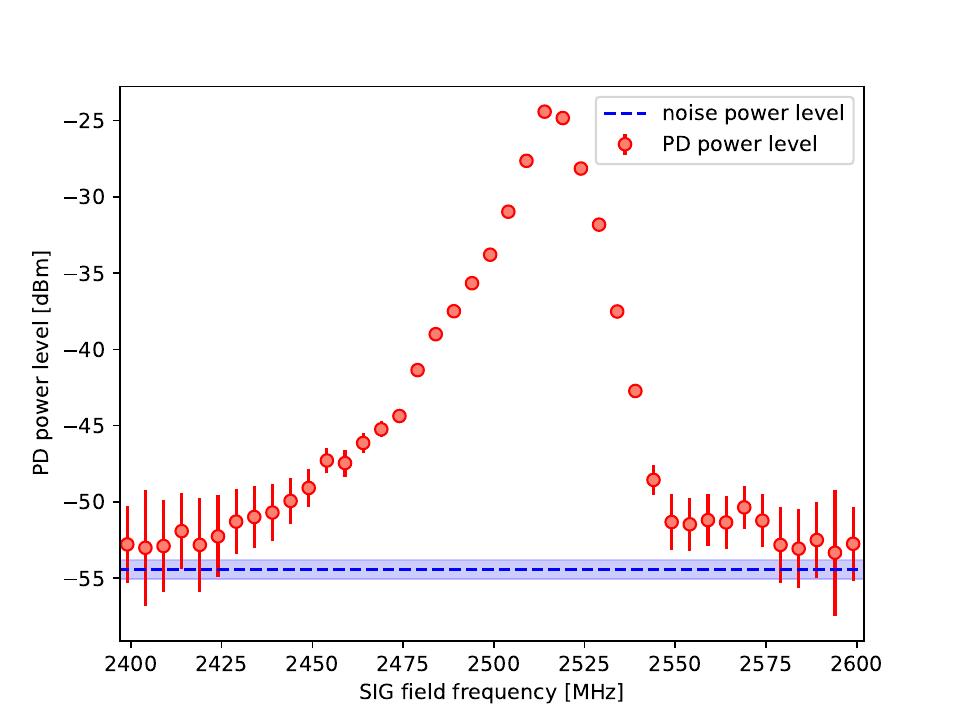}
\caption{Atomic response power level as a function of the $\mathsf{SIG}$ field frequency together with the shot-noise power level. The atomic response was measured for different frequencies of the $\mathsf{SIG}$ and $\mathsf{DRS}$ fields with a value of $f_{\mathsf{OPT}}$ being fixed and equal to $f_{\mathsf{OPT}}=\SI{5}{\MHz}$. The blue area around the shot-noise power level corresponds to the standard error of the calculated value.}
\label{fig:band}
\end{figure}

In the second case, we consider the frequency changes for both $\mathsf{SIG}$ and $\mathsf{CPL}$ fields in such a way, that the value of $f_{\mathsf{OPT}}$ remains fixed and equal to $f_{\mathsf{OPT}}= \SI{5}{\MHz}$. The measurements are performed by detuning both fields in the frequency range spanning from $\SI{-145}{\MHz}$ to $\SI{-145}{\MHz}$ with the step of $\SI{5}{\MHz}$. For each frequency, we measure the Fourier power spectrum with all the parameters being the same as in the previous case. For each of the detunings, we gather 100 consecutive waveforms of the Fourier power spectrum of the $\mathsf{PD}$ signal with the measurement time of $\SI{0.26}{\ms}$. The signal's power level is found by taking the maximum value in the frequency range of $\SI{1}{\MHz}$ around the expected beat note frequency of $\SI{5}{\MHz}$ from the averaged Fourier power spectrum and its variance is found by taking the variance of the maximum value over each of the waveforms. The measured atomic response as a function of $\mathsf{SIG}$ field frequency together with the noise power level can be seen in Fig \ref{fig:band}. As it can be seen from the figure, the atomic response drops to the near the shot-noise power level for the $\mathsf{SIG}$ frequencies of about $f_{\mathsf{SIG}} \approx \SI{2450}{\MHz}$ and $f_{\mathsf{SIG}} \approx \SI{2550}{\MHz}$. 

\section{Summary}
In this paper, we presented the phase-sensitive detection scheme using 3 different $\mathsf{MW}$ fields, one of which drives atoms via the two-photon transition. Utilizing the phase-stability of the $\mathsf{MW}$ fields, we compared measured phase-dependent transmission spectra to the theoretical predictions acquired from solving the time-dependent Lindblad equation. We characterized the receiver in terms of sensitivity, achieving the value of $\SI{3.2\pm0.3}{\micro\volt\per\cm\per\hertz\tothe{0.5}}$, as well as its frequency response range. Similar behaviour of the Rydberg atom-based receiver was also reported in other works \cite{Meyer2018, Holloway2019}. Moreover, we also measured the atomic response in case of constant $f_{\mathsf{OPT}} = \SI{5}{\MHz}$ with changing frequencies of $\mathsf{SIG}$ and $\mathsf{DRS}$ fields, for which we observed drop to the noise level at the frequencies $f_{\mathsf{SIG}} \approx \SI{2450}{\MHz}$ and $f_{\mathsf{SIG}} \approx \SI{2550}{\MHz}$.

As it can be noted on the figures depicting comparisons between numerical predictions and measured data, the fit is not perfect, and significant deviations can be observed. As mentioned earlier, such behavior is expected and can be attributed to the splitting of the degeneracy of the Rydberg state due to the strong $\mathsf{MW}$ fields. To resolve such a problem, it was already shown that one can split the degenerate structure by applying the magnetic field and introduce the Zeeman shift to the atomic ensemble \cite{Schlossberger2024}. Additionally, a similar effect could be achieved by introducing the AC-Stark shifts. Moreover, both of methods of better resolving the energy structure of Rydberg states could help better match theoretical predictions with the measurements and, in effect, allow for better understanding of closed-loop scheme dynamics.

\paragraph{Funding.} Narodowe Centrum Nauki (2021/43/D/ST2/03114); European Funds for Smart Economy (FENG.02.01-IP.05-0017/23); Foundation for Polish Science;

\paragraph{Acknowledgments.} 
This research was funded in whole  or in part by the National Science Centre, Poland, grant No. 2021/43/D/ST2/03114. The ,,Quantum Optical Technologies'' (FENG.02.01-IP.05-0017/23) project is carried out within the Measure 2.1 International Research Agendas programme of the Foundation for Polish Science, co-financed by the European Union under the European Funds for Smart Economy 2021-2027 (FENG). We thank B. Kasza for theoretical support and S. Borówka for support and discussions.

\paragraph{Disclosures.} 
The authors declare no conflicts of interest.

\paragraph{Data availability.} Data has been deposited at Harvard Dataverse \cite{Nowosielski}.

\bibliography{refs}

\begin{thebibliography}{10}
\newcommand{\enquote}[1]{``#1''}

\bibitem{Fancher2021}
C.~T. Fancher, D.~R. Scherer, M.~C.~S. John, and B.~L.~S. Marlow, \enquote{Rydberg atom electric field sensors for communications and sensing,} {\protect\JournalTitle{IEEE Transactions on Quantum Engineering}} \textbf{2}, 1–13 (2021).

\bibitem{SantamariaBotello2022}
G.~Santamaria-Botello, S.~Verploegh, E.~Bottomley, and Z.~Popovic, \enquote{Comparison of noise temperature of rydberg-atom and electronic microwave receivers,}  (2022).

\bibitem{Meyer2021}
D.~H. Meyer, P.~D. Kunz, and K.~C. Cox, \enquote{Waveguide-{Coupled} {Rydberg} {Spectrum} {Analyzer} from 0 to 20 {GHz},} {\protect\JournalTitle{Physical Review Applied}} \textbf{15}, 014053 (2021).

\bibitem{Cui2023}
Y.~Cui, F.-D. Jia, J.-H. Hao, \emph{et~al.}, \enquote{Extending bandwidth sensitivity of {Rydberg}-atom-based microwave electrometry using an auxiliary microwave field,} {\protect\JournalTitle{Physical Review A}} \textbf{107}, 043102 (2023).

\bibitem{Simons2018}
M.~T. Simons, J.~A. Gordon, and C.~L. Holloway, \enquote{Fiber-coupled vapor cell for a portable rydberg atom-based radio frequency electric field sensor,} {\protect\JournalTitle{Applied Optics}} \textbf{57}, 6456 (2018).

\bibitem{Mao2023}
R.~Mao, Y.~Lin, K.~Yang, \emph{et~al.}, \enquote{A high-efficiency fiber-coupled rydberg-atom integrated probe and its imaging applications,} {\protect\JournalTitle{IEEE Antennas and Wireless Propagation Letters}} \textbf{22}, 352–356 (2023).

\bibitem{Zhao2023}
R.~Zhao, M.~Feng, J.~Zhu, \emph{et~al.}, \enquote{Toward the measurement of microwave electric field using cesium vapor mems cell,} {\protect\JournalTitle{IEEE Electron Device Letters}} \textbf{44}, 2031–2034 (2023).

\bibitem{Sedlacek2012}
J.~A. Sedlacek, A.~Schwettmann, H.~Kübler, \emph{et~al.}, \enquote{Microwave electrometry with {Rydberg} atoms in a vapour cell using bright atomic resonances,} {\protect\JournalTitle{Nature Physics}} \textbf{8}, 819--824 (2012).

\bibitem{Osterwalder1999}
A.~Osterwalder and F.~Merkt, \enquote{Using {High} {Rydberg} {States} as {Electric} {Field} {Sensors},} {\protect\JournalTitle{Physical Review Letters}} \textbf{82}, 1831--1834 (1999).

\bibitem{Borowka2022}
S.~Borówka, U.~Pylypenko, M.~Mazelanik, and M.~Parniak, \enquote{Sensitivity of a {Rydberg}-atom receiver to frequency and amplitude modulation of microwaves,} {\protect\JournalTitle{Applied Optics}} \textbf{61}, 8806--8812 (2022).

\bibitem{Liu2021}
X.~Liu, F.~Jia, H.~Zhang, \emph{et~al.}, \enquote{Using amplitude modulation of the microwave field to improve the sensitivity of rydberg-atom based microwave electrometry,} {\protect\JournalTitle{AIP Advances}} \textbf{11}, 085127 (2021).

\bibitem{Simons2019}
M.~T. Simons, A.~H. Haddab, J.~A. Gordon, and C.~L. Holloway, \enquote{A {Rydberg} atom-based mixer: {Measuring} the phase of a radio frequency wave,} {\protect\JournalTitle{Applied Physics Letters}} \textbf{114}, 114101 (2019).

\bibitem{Gordon2019}
J.~A. Gordon, M.~T. Simons, A.~H. Haddab, and C.~L. Holloway, \enquote{Weak electric-field detection with sub-1 {Hz} resolution at radio frequencies using a {Rydberg} atom-based mixer,} {\protect\JournalTitle{AIP Advances}} \textbf{9}, 045030 (2019).

\bibitem{Jing2020}
M.~Jing, Y.~Hu, J.~Ma, \emph{et~al.}, \enquote{Atomic superheterodyne receiver based on microwave-dressed rydberg spectroscopy,} {\protect\JournalTitle{Nature Physics}} \textbf{16}, 911–915 (2020).

\bibitem{Borowka2024a}
S.~Borówka, U.~Pylypenko, M.~Mazelanik, and M.~Parniak, \enquote{Continuous wideband microwave-to-optical converter based on room-temperature {Rydberg} atoms,} {\protect\JournalTitle{Nature Photonics}} \textbf{18}, 32--38 (2024).

\bibitem{Han2018}
J.~Han, T.~Vogt, C.~Gross, \emph{et~al.}, \enquote{Coherent microwave-to-optical conversion via six-wave mixing in rydberg atoms,} {\protect\JournalTitle{Physical Review Letters}} \textbf{120}, 093201 (2018).

\bibitem{Vogt2019}
T.~Vogt, C.~Gross, J.~Han, \emph{et~al.}, \enquote{Efficient microwave-to-optical conversion using rydberg atoms,} {\protect\JournalTitle{Physical Review A}} \textbf{99}, 023832 (2019).

\bibitem{Robinson2021}
A.~K. Robinson, N.~Prajapati, D.~Senic, \emph{et~al.}, \enquote{{Determining the angle-of-arrival of a radio-frequency source with a Rydberg atom-based sensor},} {\protect\JournalTitle{Applied Physics Letters}} \textbf{118}, 114001 (2021).

\bibitem{Sedlacek2013}
J.~A. Sedlacek, A.~Schwettmann, H.~Kübler, and J.~P. Shaffer, \enquote{Atom-{Based} {Vector} {Microwave} {Electrometry} {Using} {Rubidium} {Rydberg} {Atoms} in a {Vapor} {Cell},} {\protect\JournalTitle{Physical Review Letters}} \textbf{111}, 063001 (2013).

\bibitem{Anderson2018}
D.~A. Anderson, E.~G. Paradis, and G.~Raithel, \enquote{A vapor-cell atomic sensor for radio-frequency field detection using a polarization-selective field enhancement resonator,} {\protect\JournalTitle{Applied Physics Letters}} \textbf{113}, 073501 (2018).

\bibitem{Meyer2018}
D.~H. Meyer, K.~C. Cox, F.~K. Fatemi, and P.~D. Kunz, \enquote{{Digital communication with Rydberg atoms and amplitude-modulated microwave fields},} {\protect\JournalTitle{Applied Physics Letters}} \textbf{112}, 211108 (2018).

\bibitem{Song2019}
Z.~Song, H.~Liu, X.~Liu, \emph{et~al.}, \enquote{Rydberg-atom-based digital communication using a continuously tunable radio-frequency carrier,} {\protect\JournalTitle{Optics Express}} \textbf{27}, 8848 (2019).

\bibitem{Morigi2002}
G.~Morigi, S.~Franke-Arnold, and G.-L. Oppo, \enquote{Phase-dependent interaction in a four-level atomic configuration,} {\protect\JournalTitle{Physical Review A}} \textbf{66}, 053409 (2002).

\bibitem{Kosachiov1992}
D.~V. Kosachiov, B.~G. Matisov, and Y.~V. Rozhdestvensky, \enquote{Coherent phenomena in multilevel systems with closed interaction contour,} {\protect\JournalTitle{Journal of Physics B: Atomic, Molecular and Optical Physics}} \textbf{25}, 2473 (1992).

\bibitem{Buckle1986}
S.~Buckle, S.~Barnett, P.~Knight, \emph{et~al.}, \enquote{Atomic {Interferometers},} {\protect\JournalTitle{Optica Acta: International Journal of Optics}} \textbf{33}, 1129--1140 (1986).

\bibitem{Berweger2023a}
S.~Berweger, A.~B. Artusio-Glimpse, A.~P. Rotunno, \emph{et~al.}, \enquote{Closed-loop quantum interferometry for phase-resolved {Rydberg}-atom field sensing,} {\protect\JournalTitle{Physical Review Applied}} \textbf{20}, 054009 (2023).

\bibitem{Borowka2024}
S.~Borówka, M.~Mazelanik, W.~Wasilewski, and M.~Parniak, \enquote{Optically-biased {Rydberg} microwave receiver enabled by hybrid nonlinear interferometry,} Tech. rep. (2024). ArXiv:2403.05310 [physics, physics:quant-ph] type: article.

\bibitem{Anderson2022}
D.~Anderson, R.~Sapiro, L.~Gonçalves, \emph{et~al.}, \enquote{Optical {Radio}-{Frequency} {Phase} {Measurement} {With} an {Internal}-{State} {Rydberg} {Atom} {Interferometer},} {\protect\JournalTitle{Physical Review Applied}} \textbf{17}, 044020 (2022).

\bibitem{Shylla2018}
D.~Shylla, E.~O. Nyakang’o, and K.~Pandey, \enquote{Highly sensitive atomic based {MW} interferometry,} {\protect\JournalTitle{Scientific Reports}} \textbf{8}, 8692 (2018).

\bibitem{Fleischhauer2005}
M.~Fleischhauer, A.~Imamoglu, and J.~P. Marangos, \enquote{Electromagnetically induced transparency: {Optics} in coherent media,} {\protect\JournalTitle{Reviews of Modern Physics}} \textbf{77}, 633--673 (2005).

\bibitem{Carr:12}
C.~Carr, M.~Tanasittikosol, A.~Sargsyan, \emph{et~al.}, \enquote{Three-photon electromagnetically induced transparency using rydberg states,} {\protect\JournalTitle{Opt. Lett.}} \textbf{37}, 3858--3860 (2012).

\bibitem{Kasza2024}
B.~Kasza, S.~Bor\'owka, W.~Wasilewski, and M.~Parniak, \enquote{Atomic-optical interferometry in fractured loops: A general solution for rydberg radio-frequency receivers,} {\protect\JournalTitle{Phys. Rev. A}} \textbf{111}, 053718 (2025).

\bibitem{Nowosielski2024}
J.~Nowosielski, M.~Jastrzębski, P.~Halavach, \emph{et~al.}, \enquote{Warm {Rydberg} atom-based quadrature amplitude-modulated receiver,} {\protect\JournalTitle{Optics Express}} \textbf{32}, 30027--30039 (2024).

\bibitem{Robertson2021}
E.~J. Robertson, N.~Šibalić, R.~M. Potvliege, and M.~P.~A. Jones, \enquote{{ARC} 3.0: {An} expanded {Python} toolbox for atomic physics calculations,} {\protect\JournalTitle{Computer Physics Communications}} \textbf{261}, 107814 (2021).

\bibitem{Holloway2019}
C.~L. Holloway, M.~T. Simons, J.~A. Gordon, and D.~Novotny, \enquote{Detecting and {Receiving} {Phase}-{Modulated} {Signals} {With} a {Rydberg} {Atom}-{Based} {Receiver},} {\protect\JournalTitle{IEEE Antennas and Wireless Propagation Letters}} \textbf{18}, 1853--1857 (2019).

\bibitem{Schlossberger2024}
N.~Schlossberger, A.~P. Rotunno, A.~B. Artusio-Glimpse, \emph{et~al.}, \enquote{Zeeman-resolved autler-townes splitting in rydberg atoms with tunable resonances and a single transition dipole moment,} {\protect\JournalTitle{Physical Review A}} \textbf{109}, L021702 (2024).

\bibitem{Nowosielski}
J.~Nowosielski, M.~Mazelanik, W.~Wasilewski, and M.~Parniak, \enquote{Replication {Data} for: {Superheterodyne} {Rydberg} {S}-band receiver with a multi-tone local oscillator based on an atomic transition loop,}  (2025). Type: dataset.

\end{thebibliography}

\end{document}